**Ripple formation on silver after irradiation with radially polarised ultrashort-pulsed lasers**

George D.Tsibidis,[1,a)] and Emmanuel Stratakis,[1,2]

[1]*Institute of Electronic Structure and Laser (IESL), Foundation for Research and Technology (FORTH), N. Plastira 100, Vassilika Vouton, 70013, Heraklion, Crete, Greece*
[2]*Materials Science and Technology Department, University of Crete, 71003 Heraklion, Greece*

[a)] Electronic mail: tsibidis@iesl.forth.gr

We report on the morphological effects induced by the inhomogeneous absorption of cylindrically polarized femtosecond laser irradiation of silver (Ag) in sub-ablation conditions. A theoretical prediction of the role of surface plasmon excitation and thermal effects in the production of self-formed periodic ripples structures is evaluated. To this end, a combined hydrodynamical and thermoelastic model is presented to account for the influence of temperature-related lattice movements in laser beam conditions that are sufficient to produce material melting. Results indicate that material displacements due to hydrodynamics are substantially larger than strain-related movements which also emphasises the predominant role of fluid transport in surface modification. Moreover, theoretical simulations highlight the influence of the polarisation state in the size of the ripple periodicity for a specialized case of cylindrically polarized beams, the radially polarized beams. Results show that the ripple periodicity is larger than if linearly polarized beams are used. This is the opposite trend to the behaviour for materials with decreasing electron-phonon coupling constant *g* with increasing electron temperature which highlight the significant role of *g*.

**\*** E-mail: tsibidis@iesl.forth.gr

## I. INTRODUCTION

Material processing with ultra-short pulsed lasers has received considerable attention over the past decades due to its important technological applications, in particular in industry and medicine [1-9]. One type of surface modification, the so-called laser-induced periodic surface structures (LIPSS) on solids have been studied extensively. Previous theoretical approaches or experimental observations related to the formation mechanism of these structures were performed in submelting [10] or ablation conditions [11-17]. Various mechanisms have been proposed to account for the development of these periodic structures: interference of the incident wave with an induced scattered wave [12, 14, 17], or with a surface plasmon wave (SPW) [13, 16, 18-21], or due to self-organisation mechanisms [22]. Most of these works consider irradiation with linearly polarized beams.

On the other hand, cylindrical vector beams (CVB) have recently gained remarkable attention as the symmetry of the polarization enables new processing strategies [23] with applications in various fields including microscopy, lithography [24], electron acceleration [25], material processing [23, 26, 27] and optical trapping [28]. The cylindrical vector polarization states have been the topic of numerous theoretical and experimental investigations [26, 29-32] and they are also characterized by a great potential to provide a plethora of complex structures and biomimetic surfaces [33] with a wide range of applications [34].

Recently, the mechanism of LIPSS formation on transition metals (i.e. nickel) upon femtosecond irradiation with CVB was explored systematically and the comparison with results derived by linearly polarized beams were highlighted [35]. While a combination of electrodynamic (i.e. related to surface plasmon excitation) and temperature gradient related effects were proposed to explain the periodically modulated surface for nickel which is a material characterized with strong electron-phonon coupling constant $g$, it is equally important to investigate whether metals with relatively smaller $g$ also demonstrate similarly pronounced morphologies. This is due to the fact that the magnitude of $g$ determines the efficiency/speed with which electron energy is transferred to the lattice system (i.e. strong coupling strength implies larger energy transfer from the hot electrons to lattice and therefore thermalisation between electron and lattice systems occurs faster). It is known that smaller electron-phonon coupling strengths are expected to lead to less pronounced ripples for linearly polarized beams for noble metals [36, 37]. Hence, it is evident that more intense beams or increase of irradiation would be necessary for enhancing the induced morphological features (i.e. height of ripples, depth of ablated region). Nevertheless, a more conclusive investigation requires a systematic analysis of the induced surface profile after irradiation with more complex laser beam profiles such as CVB. To illustrate the impact of this type of beams, we consider irradiation of silver (Ag) with one specialized case of CVB, the *radially polarized beams* (RP) [29], of moderate fluence that is sufficient to produce material melting and highlight fluid transport-related surface modification. It needs to be emphasized that RP light can be focused into a tighter spots than those resulting from spatially homogeneous polarisation by assuming a strong localized longitudinal component [29], however, in this work, tight focusing will not be considered.

In this work, we present numerical results obtained by modelling interaction of femtosecond (radially polarized) pulsed lasers with Ag. As explained above, it is assumed that the laser beam conditions are sufficient to melt, partially, the irradiated material. Hence, the produced surface modification of the material is dictated predominantly by fluid dynamics and secondly by induced strains in regions where energy is not sufficiently strong [10]. It is beyond the scope of the present work to explore more complex mechanisms where spallation or mass removal (i.e. ablation) effects are involved. Furthermore, we investigate the mechanisms that lead to ripple formation with increasing number of pulses (*NP*) at constant fluence. To provide a complete and rigorous description, all thermophysical parameters, material optical properties as well as the electron-phonon coupling constants are assumed to be temperature-dependent [38]. A comparison of ripple periodicity variation with increasing number of pulses for different polarization states (i.e. linearly *vs.* radially polarized beams) is also demonstrated.

## II. THEORETICAL MODEL

To understand the physical mechanism that accounts for the surface modification upon irradiation of metals with femtosecond (fs) pulsed lasers, it is important to perform a multiscale (on all time-scales) modelling of the processes that describe laser beam energy absorption and response of the material. Therefore, the theoretical model that is presented should comprise the following components: (i) a term that describes energy absorption (ii) a term that describes electron excitation, (iii) a heat transfer component that accounts for electron-lattice thermalisation through particle dynamics and heat conduction and carrier-phonon coupling, and (iv) a hydrodynamics component that describes fluid dynamics and re-solidification process (in areas where a phase transition occurs) and a component that accounts for the thermomechanical response of the material (where elastoplastic effects dominate) [35]. The following table (Table 1) provides a picture of the typical processes and the associated time-scales [39]. In principle, the processes start after some fs, they continue to mechanisms that complete after some picoseconds (ps) while others require more time and they last up to the nanosecond (ns) regime. Table 1 illustrates a typical sketch of the physical processes that may also vary with the laser beam pulse duration and strength. Other pathways (i.e. ablation, spallation, etc.) are also possible, however, the laser beam conditions used in this work are not sufficient to produce such effects. In the rest of this section, the components that constitute the theoretical framework are described [39].

**Table 1** Typical physical processes following irradiation of metals with fs pulses

| Process | Time scale |
|---|---|
| Laser beam absorption | fs |
| Electron excitation | fs |
| Electron-lattice heating | ps |
| Phase transition (melting) | ps to ns |
| Thermomechanical effects | ps to ns |
| Surface modification | ns |

### A. Polarisation of the laser beam

Irradiation of the material with a radially polarised laser beam is assumed in which the electric field of the beam is typically expressed as the superposition of orthogonally Hermite-Gauss $HG_{01}$ and $HG_{10}$ modes [29]

$$\vec{E}_r = HG_{10}\hat{x} + HG_{01}\hat{y} \tag{1}$$

where $\vec{E}_r$ denotes radial polarization and $\hat{x}$, $\hat{y}$ are the unit vectors along the *x*- and *y*-axis (Cartesian system), respectively (Fig.1). The spatial profile of the linearly polarized beam (LP) in Fig.1a is Gaussian and it is normalized to 1. Based on Eq. (1), the radially polarized beam has a maximum intensity (corresponding to the peak fluence) at $r=R_0$ with value $\sim (2\sqrt{2})^2 e^{-1}/4 \sim 0.74$ where $R_0$ stands for the irradiation spot radius ($R_0=7\mu m$ in Fig.1), similar to that used in recent works [40, 41]. In both types of polarization, the same total laser beam energy is assumed. The spatial profile of LP and RP along the $X''$-axis that denotes the radial distance from the origin as well as the two-dimensional intensity distribution of the two types of beams are illustrated in Fig.1b,c.

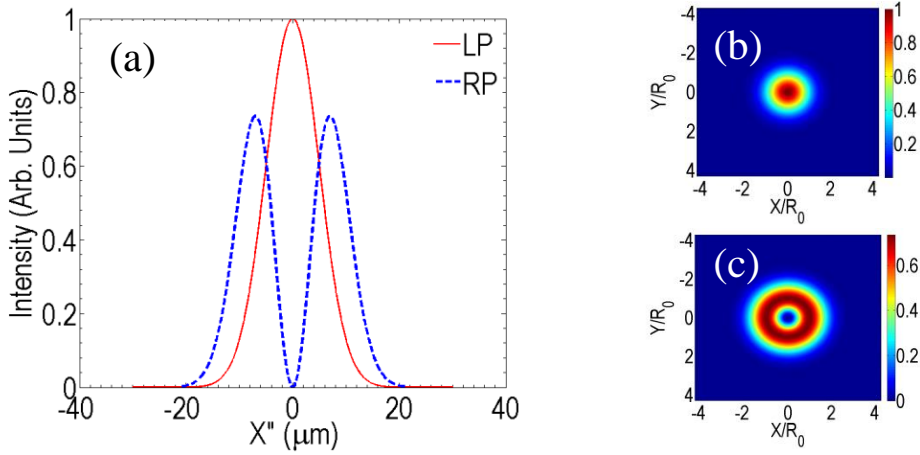

**Fig. 1** (Color Online) (a) Spatial intensity distribution of Radially (RP) and Linearly (LP) polarized beam along the $X''$-axis that denotes the radial distance from the origin (along the direction of the unit vector $\hat{r}$, $\hat{r}=(\hat{x}+\hat{y})/\sqrt{2}$), (b) Two-dimensional spatial profile of LP, (c) Two-dimensional spatial profile of RP.

## B. Energy absorption, electron excitation and electron-lattice heating

The employment of the Two-Temperature Model (TTM) constitutes the standard theoretical approach to investigate laser-matter interaction which assumes an instantaneous electron excitation [42]. Although, modified versions of TTM have been also introduced to describe electron-phonon relaxation processes for extremely very short pulses (i.e. smaller than 50fs) where an electron thermalisation process is technically incorporated [43-46], the value of the laser beam pulse duration in the simulations (~170fs) is sufficiently high to ensure that the classical TTM describes efficiently electron and lattice temperature dynamics. The TTM is characterised by the following set of equations that describe energy transfer between the electron and lattice subsystems

$$C_e \frac{\partial T_e}{\partial t} = \vec{\nabla} \cdot (\kappa_e \vec{\nabla} T_e) - g(T_e - T_L) + S(\vec{r}, t)$$

(2)

$$C_L \frac{\partial T_L}{\partial t} = \vec{\nabla} \cdot (\kappa_L \vec{\nabla} T_L) + g(T_e - T_L)$$

where $T_e$ and $T_L$ are the electron and lattice temperatures, $k_e$ and $k_L$ stand for the electron and lattice heat conductivities, $C_e$ and $C_L$ correspond to the electron and lattice heat capacities, and $g$ is the electron-phonon coupling strength. The heat source $S$ that corresponds to the laser heating is modelled assuming a Gaussian temporal profile and it has the following form

$$S(x,y,z,t) = \frac{\sqrt{4\log 2}}{\sqrt{\pi}\tau_p} E_p \frac{(1-R(x,y,z=0,t))}{1/\alpha(x,y,z,t)+\lambda_{ball}} \exp\left(-4\log 2\left(\frac{t-3\tau_p}{\tau_p}\right)^2\right) \int_0^z \frac{1}{2}|\vec{E}_r(x,y,z',t)|^2 \exp\left(-\frac{z'}{1/\alpha(x,y,z',t)+\lambda_{ball}}\right)dz' \quad (3)$$

where $\tau_p$ is the pulse duration ($\tau_p$ =170fs), $R_0$=15μm, $E_p$ is the *peak fluence for LP* ($E_p$ =1.4J/cm$^2$), and $\lambda_{ball}$ (=53.7nm [42, 47, 48]) is the ballistic electron penetration depth. The computation of the absorption coefficient $\alpha$ and reflectivity $R$ is performed through the computation of the dielectric constant by assuming the extended Lorentz-Drude model with five Lorentzian terms based on the analysis of Rakic *et al.* [49]

$$\varepsilon(\omega_L) = 1 - \frac{f_0 \omega_p^2}{\omega_L^2 - i\gamma\omega_L} + \sum_{j=1}^{k=5} \frac{f_j \omega_p^2}{\omega_j^2 - \omega_L^2 + i\omega_L \Gamma_j} \quad (4)$$

where $\omega_L$ is the laser frequency (=2.3562 × 10$^{15}$ rad/s for 800nm that corresponds to photon energy equal to 1.55eV), and $\sqrt{f_0}\omega_p$ is the plasma frequency associated with oscillator strength $f_0$ and damping constant $\gamma$ (that equals the reciprocal of electron relaxation time). The interband part of the dielectric constant (third term in Eq.(4)) assumes five oscillators with frequency $\omega_j$, strength $f_j$, and lifetime $1/\Gamma_j$. Values for the aforementioned parameters are given in Rakic *et al.*[49] (i.e. $f_0$=0.845, h$\omega_p$=9.01eV, $\omega_1$=0.816eV, $\omega_2$=4.481, $\omega_3$=8.185, $\omega_4$=9.083, $\omega_5$=20.29, $\Gamma_1$=3.886eV, $\Gamma_2$=0.452, $\Gamma_3$=0.065, $\Gamma_4$=0.916, $\Gamma_5$=2.419, $f_1$=0.065, $f_2$=0.124, $f_3$=0.011, $f_4$=0.84, $f_5$=5.646). The temporal dependence of the dielectric constant and the optical parameters come from the electron relaxation time which is the sum of the electron-electron and electron-phonon collision rates, $A(T_e)^2$ and $BT_L$, respectively [50, 51]. Hence, the dynamic character of the optical properties is incorporated in the simulations through the temporal change of $\gamma$. The value of the electron conductivity $k_e$ is given by the following expression

$$k_e = \frac{\left((\theta_e)^2 + 0.16\right)^{5/4}\left((\theta_e)^2 + 0.44\right)\theta_e}{\left((\theta_e)^2 + 0.092\right)^{1/2}\left((\theta_e)^2 + \eta\theta_L\right)^{5/4}} \chi \quad (5)$$

$$\theta_e = \frac{T_e}{T_F}, \theta_L = \frac{T_L}{T_F}$$

where, in the case of silver, the parameters that appear in the expression are the Fermi temperature [44] $T_F$=6.38×10$^4$K, and $\chi$=500WK$^{-1}$sec$^{-1}$, $\eta$=0.1715. On the other hand, $g$ and $C_e$ are evaluated through a fitting procedure based on results derived by calculating the electron density of states (DOS) [38]. Finally, the lattice thermal conductivity $k_L$ is taken as 1% of the thermal conductivity of bulk metal since the mechanism of heat conduction in metal is mainly due to electrons.

## C. Phase transition and thermomechanical response

In principle, any surface or structural modification requires lattice displacements or deformation due to thermal effects. To this end, it is important to analyse thoroughly the evolution of the lattice temperature of the material and explore the underlying physical processes that lead to material deformation. In this work, emphasis will be given on processes that induce phase transitions (i.e. melting) and/or large strain effects. Therefore, the laser conditions used in the simulations are appropriately selected to highlight the role of hydrodynamics and elasto-plastic processes in surface modification. To this end, a phase transition from solid-to-liquid is incorporated in the modelling framework while an investigation of strain generation determines whether plastic effects can also influence the final surface profile. In a previous work [16], a minimum mass removal was considered (i.e. ablation conditions) and therefore a more complex scenario was investigated (i.e. modelling of dynamics of lattice sites attaining temperatures well above the boiling temperature (~2435K for silver [52]), consideration of ablation and role of recoil pressure, relaxation processes, etc). By contrast, the aim of the present work is to exclude those complex processes; the focus is restricted to conditions that induce molten material movement in a subregion of the irradiated area while, elsewhere, modification is determined by the development of stresses larger than the yield stress of the material. Due to the fact that surface modification can result from the two different and distinct processes, it is important to explore the influence of each one, separately. Although, it can be argued that a combined effect is possible in some regions, it is noted that as fluid transport related morphological modification is expected to induce substantially larger displacements, it is assumed that in those regions only hydrodynamics dictates the process. The separation of the regions where the two processes occur is based entirely on the lattice temperature value (melting occurs in the subregion where $T_L>T_{melt}$). Moreover, as the lattice temperature accounts for the type of the dominant mechanism, its evolution is required to be investigated by assuming appropriate heat diffusion equations in the two regimes.

### (i) Phase transition

To describe the heat transfer in the molten material and analyse fluid dynamics, it is assumed that the fluid behaves as an incompressible Newtonian fluid. Then, fluid dynamics is provided by the following set of equations (to account for the mass, energy, and momentum conservation, respectively) [16]

$$\vec{\nabla} \cdot \vec{u} = 0$$

$$\left(C_L^{(m)} + L_m \delta(T_L - T_{melt})\right)\frac{\partial T_L}{\partial t} + C_L^{(m)} \vec{\nabla} \cdot (\vec{u} T_L) = \vec{\nabla} \cdot (\kappa_L^{(m)} \vec{\nabla} T_L) \tag{6}$$

$$\rho_L^{(m)} \left(\frac{\partial \vec{u}}{\partial t} + \vec{u} \cdot \vec{\nabla} \vec{u})\right) = \vec{\nabla} \cdot \left(-P + \mu^{(m)}(\vec{\nabla}\vec{u}) + \mu^{(m)}(\vec{\nabla}\vec{u})^T\right)$$

where $\vec{u}$ is the fluid velocity, $\mu^{(m)}$ is the dynamic viscosity of the liquid, $\rho_L^{(m)})$ is the density of the molten material, $T_{melt}$ is the melting temperature for Ag, $L_m$ is the latent heat for phase transformation, $P$ stands for pressure, $k_L^{(m)}$ and $C_L^{(m)}$ stand for lattice heat conductivity (see Table 2) and lattice heat capacity respectively, of the molten material. The value of $k_L^{(m)}$ is equal to $\alpha_D C_L^{(m)} \rho_L^{(m)}$, where $\alpha_D = 8 \times 10^{-5} \text{m}^2/\text{sec}$ [53] stands for the thermal diffusivity of liquid Ag. The term that contains the delta function has been presented to provide a smooth transition from the solid-to liquid phase and describe efficiently the resolidification process [16, 54]. Phase transformation directly influences the heat capacity of the material. Therefore, $L_m \delta(T_L - T_{melt})$ is considered as a correction to the heat capacity on the solid-liquid interface to provide a smooth transition for the heat capacity between the two phases. As the movement of the $T_{melt}$ isothermal (inside the volume of the irradiated material) is used to determine the resolification process, smoothness of the thermophyical parameters for temperatures above and below this temperature is achieved by a suitable representation of the delta function (i.e. Gaussian) of the form $\delta(T_L - T_{melt}) = \frac{1}{\sqrt{2\pi}\Delta} e^{-\left[\frac{(T_L - T_{melt})^2}{2\Delta^2}\right]}$ where $\Delta$ is in the range of 10-100K depending on the temperature gradient [16, 55].

### (ii) Thermomechanical response

To describe the thermomechanical response of the material in regions where strain-related effects dominate, the following dynamic equations of elasticity are used

$$\rho_L^{(s)} \frac{\partial^2 V_i}{\partial t^2} = \sum_{j=1}^{3} \frac{\partial \sigma_{ji}}{\partial x^j}$$

$$\sigma_{ij} = 2\mu \varepsilon_{ij} + \lambda \sum_{k=1}^{3} \varepsilon_{kk} \delta_{ij} - \delta_{ij}(3\lambda + 2\mu)\alpha'(T_L - T_0) \tag{7}$$

$$\varepsilon_{ij} = 1/2 \left(\frac{\partial V_i}{\partial x^j} + \frac{\partial V_j}{\partial x^i}\right)$$

where $\sigma_{ij}$ and $\varepsilon_{ij}$ stand for stress and strains, $\lambda$ ($\equiv \frac{Ev}{(1+v)(1-2v)}$) and $\mu$ ($\equiv \frac{E}{2(1+v)}$) are the Lame's coefficients, while $E$ and $v$ are the Young's modulus and Poisson ratios, respectively. On the other hand, $V_i$ are the displacements along the $x^i$ direction ($i=1,2,3$), $\rho_L^{(s)}$ is the density of (solid) Ag and $a'$ corresponds to the thermal expansion coefficient. Finally, $\delta_{ij}$ stands for Kronecker delta [56]. Furthermore, to describe more accurately the electron-phonon relaxation process and the thermally induced strains and displacements, an additional term is incorporated in the equation that provides the lattice temperature evolution to account for the strain-related temperature variation (second equation in Eq.(7)) [10, 57]

$$C_e \frac{\partial T_e}{\partial t} = \vec{\nabla} \bullet (\kappa_e \vec{\nabla} T_e) - g(T_e - T_L) + S(\vec{r}, t)$$
$$C_L \frac{\partial T_L}{\partial t} = \vec{\nabla} \bullet (\kappa_L \vec{\nabla} T_L) + g(T_e - T_L) - (3\lambda + 2\mu)\alpha' T_L \sum_{i=1}^{3} \dot{\varepsilon}_{ii} \qquad (8)$$

### E. Surface plasmon excitation

It is known that irradiation of metals with fs pulses gives rise to excitation of surface plasmon polariton (SPP) [13, 36]. The predominant scenario of ripple formation attributes the periodic surface modification to the interference of a Surface Plasmon wave with the incident electromagnetic field of the laser beam [13, 16, 18-21], and therefore, elaboration of the conditions that allow coupling of the two waves is necessary. Theoretical simulations indicate that no ripples are formed for a single pulse (*NP*=1) irradiation. In fact, SPP excitation and coupling with the incident light is not possible on a flat surface as the laser and SPP dispersion curves do not meet [16, 58-60]. Nevertheless, even for a single shot [61, 62], SPP can be excited at the presence of wavelength size defects or by the self-diffraction of the incident beam on a microscale short-lived optical inhomogeneity of the excitation region [63]. On the other hand, it turns out that the presence of gratings with period of the size of the periodicity of the SPP [64] allow ripple formation as the appropriate condition for surface plasmon excitation is satisfied [60]. Furthermore, corrugated surfaces also allow SPP excitation [16, 60, 65] and therefore, (especially in case of repetitive irradiation) correlation of morphological characteristics of the irradiated zone (depth, *Δ* and grating periodicity, *Λ*) with the magnitude of the longitudinal wavevector of the SPP requires a systematic analysis of the propagation of the respective electromagnetic field.

To include the excitation of SPP and its role in the surface modification process in the theoretical model, the electric field of the SPP has to be introduced. The interference of the incident beam with the SPP will be presented by the replacement of the electric field $\vec{E}_r$ in Eq.3 by the resultant of the field of the incident beam and that of the SPP. The interference leads to a periodic energy deposition which upon relaxation induces a spatially periodic response of the material including the lattice temperature [16]. As a result, a subsequent phase transition or thermally-generated elastic/plastic deformations will also be characterised by a spatially periodic variation leading eventually to the formation of periodic structures upon freezing.

To provide a detailed analysis of correlation of the morphological features of the periodic structure, the spatio-temporal distribution of the electric field is modelled assuming a surface profile that is determined by the approximating function (where $r=(x^2+y^2)^{1/2}$)

$$\Delta(r) = D(r)\sin(2\pi r / \Lambda) \qquad (9)$$

In the case of a spatially Gaussian beam (for linearly polarized beams), $D(r) \sim exp(-r^2)$. By contrast, for a RP, $D(r)$ is determined by the form of the spatial dependence of the intensity resulting from Eq.(1). The interaction between the incident beam and the excited surface-plasmon wave is estimated assuming low modulation of the surface grating structure. The determination of the spatial distribution of the electric field and derivation of the dispersion relations comes from the solution of the Maxwell's equations by imposing the boundary condition that the tangential component of the total electric field $\vec{E}$ and normal component of $\varepsilon\vec{E}$ ($\varepsilon$ stands for the dielectric constant of Ag) should be equal [35]. To this end, we considered a 3D Cartesian coordinate system defined by $X''$-$Y''$-$z$, where $Y''$ is perpendicular to the $X''$-$z$ (or $r$-$z$ plane). The electric and magnetic fields are

$$\vec{E}^{(j)} = \begin{pmatrix} E_{X''}^{(j)} \\ 0 \\ E_z^{(j)} \end{pmatrix} e^{-i\omega t} e^{ik_{X''}^{(j)} X'' - k_z^{(j)} z}$$

$$\vec{H}^{(j)} = \begin{pmatrix} 0 \\ H_{Y''}^{(j)} \\ 0 \end{pmatrix} e^{-i\omega t} e^{ik_{X''}^{(j)} X'' - k_z^{(j)} z}$$

(10)

where the subscript $j$ (is + or -) is used to indicate that it refers to regions above (denoted as $AR$) or below (denoted as $BR$), respectively, the separating line that defines the surface morphology (Fig.2). In Eq.(10), $\omega$ is the angular frequency to be determined from the dispersion relations, $k_{X''}^{(+)} = k_{X''}^{(-)} \equiv k_{X''}$ is the component of the wavevector of the surface wave along the $X''$-axis while the $k$-vector is [66]

$$k_z^{(j)} = \left( \left( k_{X''}^{(j)} \right)^2 - \frac{\omega^2}{c^2} \varepsilon^{(j)}(\omega) \right)^{1/2}$$

(11)

In Eq.(11), $\varepsilon^{(j)}$ is the dielectric constant in regions $AR$ (it is assumed that $\varepsilon^{(A)}=1$) and $BR$. At the interface of the produced grating, the boundary conditions that require the tangential component of the electric field and the normal component of $\vec{D} = \varepsilon(\omega)\vec{E}$ are continuous. Thus, the tangential electric field component is computed by the following expression

$$E_{//}(X",z) = E_{//}^{(1)}(X",z) + E_{//}^{(2)}(X",z) = \left(E_{X"}(X",z) + \tan(\varphi) E_z(X",z)\right)\cos(\varphi) \quad (12)$$

where (see Fig.2)

$$\cos(\phi) = \sqrt{\frac{(dX")^2}{(dX")^2 + (d\Delta(X"))^2}} \quad \text{and} \quad \frac{d\Delta(X")}{dX"} = \tan(\varphi) \quad (13)$$

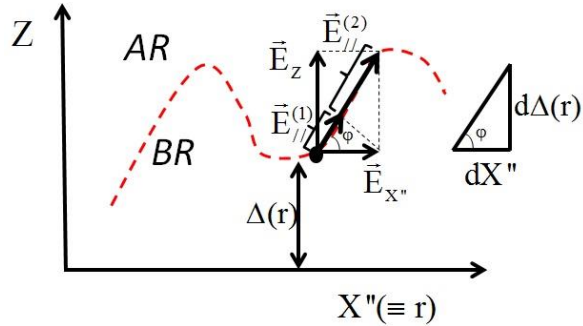

**Fig. 2** (Color Online) Tangential electric field component ($\vec{E}_{//}^{(1)} + \vec{E}_{//}^{(2)}$) on the material surface (denoted by *dashed* line).

Thus,

$$E_{//}(X",z) = \left(E_{X"}(X",z) + \frac{d\Delta(X")}{dX"} E_z(X",z)\right) \times \sqrt{\frac{(dX")^2}{(dX")^2 + (d\Delta(X"))^2}} \quad (14)$$

Similarly, the normal component of $\vec{D}$ can be computed and it is equal to

$$D_\perp(X",z) = \left(D_{X"}(X",z) - \frac{d\Delta(X")}{dX"} D_{X"}(X",z)\right) \times \sqrt{\frac{(dX")^2}{(dX")^2 + (d\Delta(X"))^2}} \quad (15)$$

The subscript symbols // and ⊥ correspond to the parallel and perpendicular component to the separating line.

A computational approach was performed to estimate the combination of grating periodicity and depth of the periodical profile that induces maximum longitudinal component of the electric field inside the ripple well. To this end, a numerical scheme is developed where Eqs.(9-15) are solved to ensure the best (optimum) SPP-laser beam coupling [13]. Results are illustrated in Fig.3. The decrease of the SPP wavelength with increasing depth (result of repetitive irradiation) has also been predicted in other materials (i.e. silicon [16] and Ni [35]) and it is a physical outcome of the need to ensure best SPP-laser coupling in a periodically corrugated profile [13].

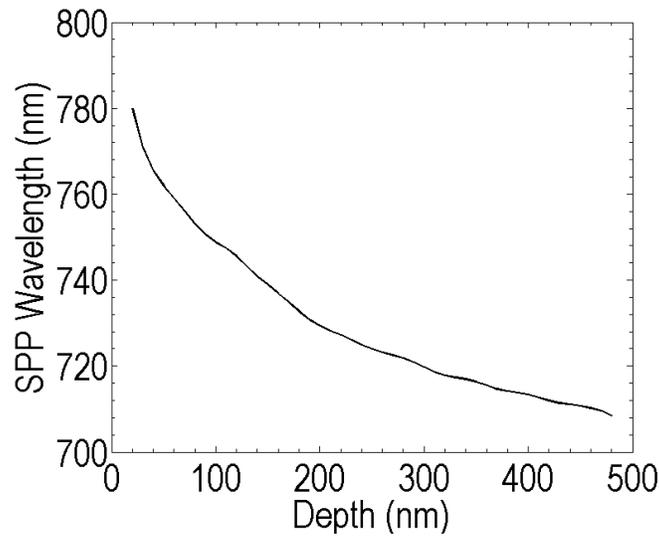

**Fig.3** Dependence of SPP wavelength on maximum grating depth of modified profile.

**F. Simulation procedure**

The parameters that are used in the numerical solution of the governing (coupled) equations Eqs. (1-8) are listed in Table 2. A finite-difference method is employed [16, 57] to solve numerically the heat transfer equations [16], phase change [16, 67] and/or dynamic elasticity equations [10]. The coupling of the various components/processes is performed in the following way: the temporal range (from the beginning of the laser irradiation to the completion of the surface modification) is discretized and, at each time point, the laser energy absorption is computed by Eqs.(1,3,4). The electron excitation and electron-relaxation process is, then, calculated by using the TTM (Eq.2) and Eq.(5). The next step is to describe the thermal effects that determine the lattice system response through the value of the lattice temperature $T_L$ which appears in Eqs.(6,7,8) (and in the thermophysical properties). Hence, a discrimination of the type of process that will follow (melting or mechanical response) will be determined by $T_L$. Plastic effects are considered when the total stress exceeds the yield stress of the material (~100MPa). The grid size taken for simulations is 2nm (vertical dimension) and 5nm (horizontal dimension) in the region where material melts while the grid size is [0.1nm×0.05nm] where thermomechanical effects dominate. Therefore, the irradiated region is split into two subregions. The temporal step is adapted so that the stability Neuman condition is satisfied [57]. As explained in previous paragraphs, the determination of the subregions where Navier-Stokes (fluid dynamics) or elasticity equations (stress-related effects) are used is based on whether the lattice temperature $T_L$ attains values larger or smaller than $T_{melt}$, respectively. On the other hand, to model the rest of the physical processes described in Section 2 (energy absorption, electron excitation and electron-lattice relaxation), although Table 1 shows that they are characterized by typical timescales, no steep or manual cut-offs are introduced. More specifically, energy absorption follows a temporal dependence and it dies off exponentially while electron excitation starts after irradiation onset. The electron relaxation time is contained in $\gamma$ (Eq.(4)) while Eqs.(2) will determine the heat transfer between the electron and lattice systems. A special attention is required, though, for the electron-lattice relaxation. More specifically, Eqs.(2) is coupled with either Eqs.(6) (or Eqs.(7)), before temperature equilibrium is reached depending on whether part of the material has undergone a phase transition (or thermomechanical effects start to occur). The distinction between the two regimes is based on the $T_L$-$T_{melt}$ value and a delta function ($\delta(T_L - T_{melt}) = \frac{1}{\sqrt{2\pi}\Delta} e^{-\left[\frac{(T_L - T_{melt})^2}{2\Delta^2}\right]}$) is used (see Section 2) to provide a smooth transition between the two subregions. After electron-lattice equilibration, the TTM is degenerated to one temperature heat conduction model (in Eq.(8), $T_e=T_L$). Although, the TTM could be used and consider technically two systems (electron and lattice) at same temperature, for the sake of simplicity and faster computation, it is assumed that the one-temperature model sufficiently describes the process. The timescale for this assumption has been set to $t$=150ps (sufficiently later than $t$ at which $T_e=T_L$ occurs). Finally, the last process, resolidification (or plasticity) is determined by using a thermal (or stress) threshold criterion. The fluid dynamics module stops upon reaching temperatures smaller than $T_{melt}$. By contrast, in regions where the solution of the elasticity equations lead to a total stress larger than the yield stress, plasticity occurs, and the operation of the thermomechanical code terminates [10].

Regarding the adequate description of fluid dynamics, the hydrodynamic equations are solved in the subregion that contains either solid or molten material. To include the "hydrodynamic" effect of the solid domain, material in the solid phase is modeled as an extremely viscous liquid ($\mu_{solid} = 10^5 \mu_{liquid}$), which results in velocity fields that are infinitesimally small. An apparent viscosity is then defined with a smooth switch function to emulate the step of viscosity at the melting temperature. A similar step function is also introduced to allow a smooth transition for the rest of the temperature-dependent quantities (i.e., heat conductivity, heat capacity, density, etc.) between the solid and liquid phases. For time-dependent flows, a common technique to solve the Navier-Stokes equations is the projection method, and the velocity and pressure fields are calculated on a staggered grid using fully implicit formulations. More specifically, the horizontal and vertical velocities are defined in the centers of the horizontal and vertical cells faces, respectively, where the pressure and temperature fields are defined in the cell centers (see description in Tsibidis et al. [16]). Similarly, all temperature-dependent quantities (i.e., viscosity, heat capacity, density, etc.) are defined in the cell centers. While second-order finite difference schemes appear to be accurate for $NP=1$, where the surface profile has not been modified substantially, finer meshes and higher-order methodologies are performed for more complex profiles [68, 69]. Furthermore, techniques that assume moving boundaries (i.e. solid-liquid interface) are employed [70]. Similarly, a finite-difference method using staggered grid is employed to solve the dynamic equations of elasticity (Eq.7) where stress, strains are computed on the cell centres while the shear stress and displacements are evaluated on the cell faces [10, 57].

**Table 2** Simulations parameters for Ag

| Parameter | Value |
|---|---|
| $\lambda_{ball}$ [nm] | 53 [48] |
| $g$ [Wm$^{-3}$K$^{-1}$] | From fitting [38] |
| $C_e$ [Jm$^{-3}$K$^{-1}$] | From fitting [38] |
| $C_L$ [$10^6$ Jm$^{-3}$ K$^{-1}$] | 2.50 [52] |
| $C_L^{(m)}$ [$10^6$ Jm$^{-3}$ K$^{-1}$] | (2.97-4.72×10$^{-4}$ $T_L$ (K$^{-1}$)) [71] |
| $A$ [$10^7$ s$^{-1}$ K$^{-2}$] | 0.932 [50, 51] |
| $B$ [$10^{11}$ s$^{-1}$ K$^{-1}$] | 1.02 [50, 51] |
| $T_{melt}$ [K] | 1235 [52] |
| $T_0$ [K] | 300 |
| $E$ [GPa] | 83 [52] |
| $N$ | 0.37 [52] |
| $\alpha'$ [$10^{-6}$K$^{-1}$] | 18.9 [52] |
| $L_m$ [kJ/kgr] | 104.57 [52] |
| $\sigma$ [N/m] | 0.92 [52] |
| $\rho_L^{(m)}$ [Kgr/m$^3$] | 10465-0.9067$T_L$ (K$^{-1}$) [72] |
| $\rho_L^{(s)}$ [Kgr/m$^3$] | 10490 [52] |
| $\mu^{(m)}$ [mPa sec] | 3.88 [52] |

The rest of the equations of the proposed model (Eqs.9-15) are related to the SPP excitation and its interference with the incident beam are used for *NP*>1 to compute the optimum SPP wavelength. The calculated SPP wavelength provides a periodic (spatial) modulation of the resultant energy that is deposited and it is subsequently incorporated in Eq.3 to describe the energy absorption and the lattice response as described in the previous paragraphs. Therefore, the inclusion of Eqs.9-15 is necessary to provide a detailed picture of the mechanism that leads to LIPSS formation.

With respect to the initial and boundary conditions, at time *t*=0, both electron and lattice temperatures are set to 300K while stress, strains and displacement components are initially zero. The same conditions are applied at the beginning of every iteration which is prescribed by *NP*. Special attention is required, though, for the accurate calculation of the laser energy absorption as the surface profile is modified after irradiation. Due to the spatially modulated profile, a ray tracing scheme is performed to calculate the energy absorption/accumulation at each point of the surface before Eqs.(3-4) are subsequently used. The top and bottom faces of the material are stress-free while elastic effects should also disappear near the edges. Non-slipping conditions (i.e. velocity field is zero) are applied on the solid-liquid interface while the condition $\mu^{(m)} \frac{\partial u_z}{\partial z} = \frac{\partial \sigma}{\partial T_s} \frac{\partial T_s}{\partial x}$ is applied on the upper and free surface of the molten material. Besides the surface tension $\sigma$, the lattice temperature of the surface, $T_{surface}$, is used [16].

## III. RESULTS

The surface modification related processes on the irradiated material are prescribed by the temperature values attained by the lattice. Therefore, the exploration of the spatio-temporal distribution of the lattice temperature profile is required to understand the complete mechanism. In Fig.4, the maximum values of the electron and lattice temperatures, $T_e$ and $T_L$, respectively, have been illustrated. Compared to other metals with stronger electron-phonon coupling constant (i.e. Ti [44]), for silver, relaxation process is delayed as a result of the smaller *g*. By contrast, the time at which equilibration between electron and lattice temperatures (~10ps) is achieved, agrees well with theoretical predictions in previous studies that suggests that time equilibration for silver lies between that for copper and gold (i.e. noble metals of comparable electron-phonon coupling strength, respectively) and therefore the expected value appears to be close to the theoretically calculated one [51, 73-77]. It is important to note, though, that in those works, the electron-phonon coupling constant and the heat capacity were approximated with well-known expressions which constitutes a different methodology from that presented in this work (based on a more accurate approach using first-principles electronic structure calculations of the electron density of states [38]). Nevertheless, the expected monotonicity (i.e. decrease of equilibration time with increase of the electron-phonon coupling constant) should still hold.

Simulations predict that the maximum $T_L$ for RP is lower than the maximum $T_L$ for LP (Fig.4). This is due to the smaller peak fluence of RP (as shown in Fig.1); it also needs to be emphasized that the maximum values occur at $X''=\pm R_0$ for RP and $X''=0$ for LP due to the different spatial distribution of the laser beam energy. The non-equilibrium state between $T_e$ and $T_L$ (after relaxation) illustrated in Fig.4a characterizes the thermal response both for LP and RP; it is explained by considering the influence of two competing mechanisms, the diffusion of electron thermal energy into deeper parts and the exchange of energy between electrons and lattice. It is evident that the former mechanism dominates which is also a characteristic of metals with a relatively weak $g$ [36] and significantly higher heat conductivity $k_e$ (see for example, Chen *et al.* [51] where typical values of $g$ for Ag, Cu, Ni are 0.31, 1, 3.6, respectively; similarly, typical values for $k_e$ for the same materials are 428, 401, and 90 J/m$^{-1}$s$^{-1}$K$^{-1}$, respectively). A similar behaviour has been reported in previous works [75, 76] where it was stated that the larger the fluence the more pronounced the discrepancy is. The discrepancy is further enhanced by the fact there is also an increased penetration due to the inclusion of the ballistic transport of the hot electrons [42]. It is evident that the material melts at $t=1.2$ps as the lattice temperature exceeds $T_{melt}$ (Fig.4). Furthermore, it is noted that for both polarization states, the lattice temperature never exceeds the boiling point of Ag. More specifically, the maximum lattice temperature for RP and LP are 1548K and 2180K, respectively, which are between the melting and boiling points. Simulations show that for all *NP* used in this work (up to *NP*=400) the lattice temperature does not exceed the boiling point (~2435K) for either of the two polarization states for $E_p = 1.4$J/cm$^2$. To obtain an estimate, though, of the ablation threshold (i.e. defined as the fluence value at which boiling occurs), a series of simulations indicated that for $E_p \sim 1.65$J/cm$^2$ the presented theoretical framework needs to be enriched with terms that account for mass removal [16].

The spatial distribution of the lattice temperature at $t=25$ps (for *NP*=1) is illustrated in Fig.5a. Lattice temperature values determine the region where lattice displacements will be dictated by hydrodynamics and surface-tension related forces (where $T_L>T_{melt}$) or mechanical effects (where $T_L<T_{melt}$). In case of phase transformation, the movement of the molten material is expected to induce two craters around the region of highest temperatures. Similar simulations have been performed for Ni [35]. By contrast, thermoelastic effects will dominate the region around $X''=0$ as $T_L<T_{melt}$ and therefore, strain-related lattice displacements will determine surface modification in the vicinity of the spot centre (certainly, thermomechanical effects occur also at distances larger than *1.5R$_0$*, however, although, possible deformation is still computed, discussion is focused on processes inside the spot). To compare the simulation results for RP with the corresponding predictions for LP, $T_L$ is illustrated at $t=25$ps for LP (Fig.5b). It is evident that the difference in the thermal response between the two polarization states is expected to be reflected on the induced morphology. In Fig.5c, the 2D surface modification for *NP*=1 is illustrated for RP. It is evident that, for a RP beam, two craters are produced followed by four humps. The humps result from mass conservation requirements (Fig.5d) as surface tension forces displace the fluid material away from each crater leading to the humps when solidification ends. By contrast, for LP, a single crater is produced as explained in previous studies (see Tsibidis *et al.* [16] and references therein).

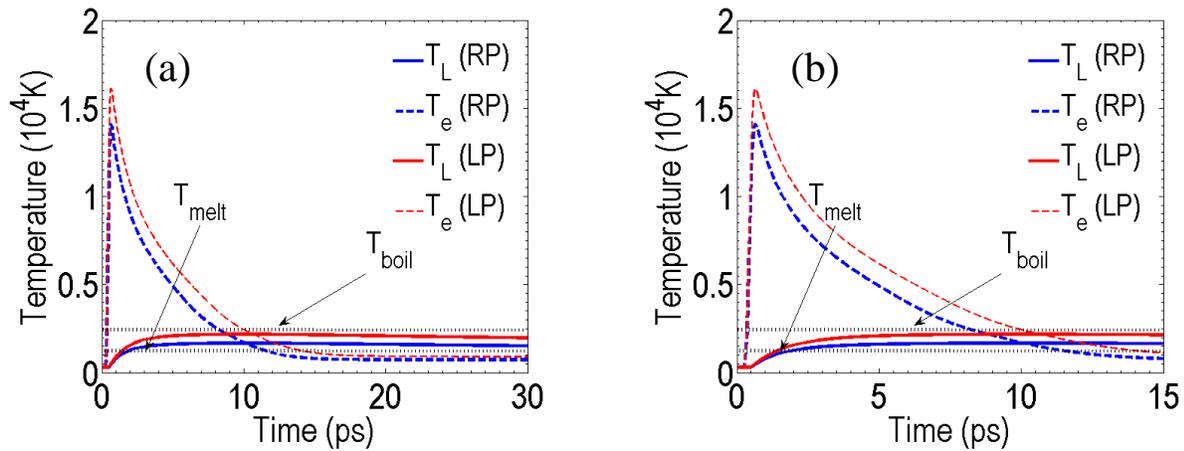

**Fig. 4** (Color Online) Evolution of maximum electron and lattice temperatures for times in the range (a) [0,30ps] and (b) [0,15ps]. The maximum values attained are above the melting and boiling temperatures indicated by the *dotted* horizontal lines (*NP*=1).

Next, simulations and analysis were performed for *NP*>1 to highlight the periodic structure formation when the material is exposed to repetitive irradiation. The spatial profile of the lattice temperature field as a result of the influence of crater formation (for *NP*=1) and the surface plasmon excitation for *NP*=2 is illustrated in Fig.6a (similar results can be obtained for *NP*>2). Due to the large laser spot area, effects are not easily identifiable and therefore, an enlarged region is illustrated. The temperature field in that region is, thus, presented in Fig.6b. The profile depression is shown in Fig.6 and it is explained by the mass displacement resulting due to fluid transport. To emphasise on the region that experiences mass displacement due to fluid transport, the illustration depicts the region below $Z=0$. The humps appear at $Z>0$ (as shown in Fig.5d).

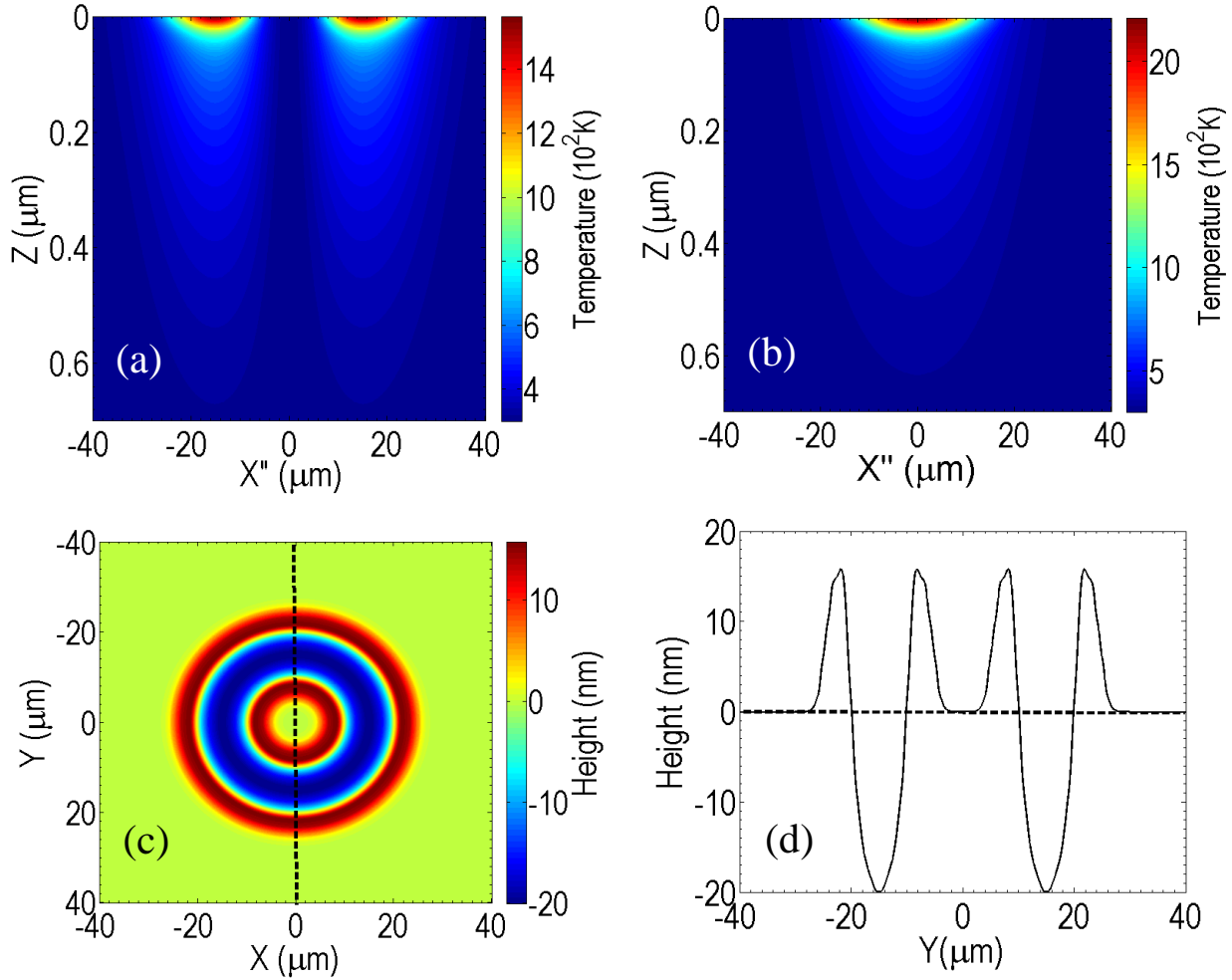

**Fig. 5** (Color Online) Lattice temperature profile at *t*=25ps (*NP*=1) for (a) RP and (b) LP. (c) Surface profile for RP. (d) Height profile along the *dashed* line in (c) (*NP*=1).

To explore the surface modification related processes through a systematic investigation of the fluid and elasticity associated movements, the focus has been restricted on the surface-tension generated movement and lattice deformations, respectively. The resolidification analysis, determined by following the evolution of the isothermal $T_{melt}$, allows calculation of the corrugation depth [20]. To discriminate on the hydrodynamics and strain generated surface modification, both horizontal and vertical displacements have been explored in regions where fluid transport ('Region 1') or elastic effects ('Region 2') dominate. More specifically, total horizontal deformation and height in Regions 1 and 2 at time $t$=0.45ns are illustrated in Figs.7-8, respectively. It is evident that both types of displacement demonstrate a spatial periodical character which originates from the interference of the surface plasmon wave with the incident beam that is subsequently reflected on the response of the material. Furthermore, results show that the morphological deformation due to surface tension forces is substantially bigger than those generated from mechanical effects. More specifically, lattice displacements due to strain effects are picometre-sized which emphasizes the predominant role of hydrodynamics in Region 1 where heat effects are substantially stronger. Due to the enhanced temperature gradient and heat conductivity along the direction of the laser beam propagation, the induced movement along the $z$-axis is anticipated to be substantially larger than the horizontal displacement. On the other hand, the predicted periodicity is closely related to the underlying physical mechanism that describes lattice response at larger timescales (~773nm/778nm in regions where phase transition/thermomechanical response dominate for $NP$=2). It needs to be emphasised that the horizontal or vertical deformation fields indicate the absolute surface profile and not relative displacements on the mass that previously occupied a particular region.

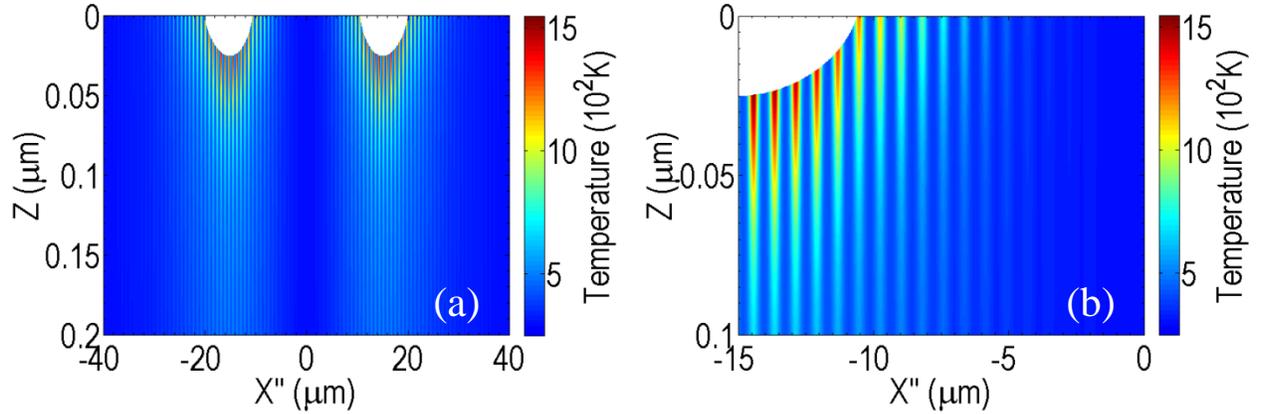

**Fig. 6** (Color Online) (a) Lattice temperature profile at $t$=25ps, (b) Lattice temperature profile at $t$=25ps (enlarged region). ($NP$=2). *White* region (air) indicates the profile depression due to mass displacement.

To derive a comprehensive and complete quantitative analysis of the morphological effects induced by radially polarized fs beams, the ripple wavelength dependence on the number of pulses has also been explored. Theoretical results illustrated in Fig.9a indicate that as the number of pulses increases, the ripple frequency increases which complies also with experimental observations for irradiation with linearly polarized beams [16]. To emphasise on the role of hydrodynamics and elasto-plastic effects in the produced frequencies, the periodicity due to (i) only SPP excitation (*solid line* in Fig.9a) and (ii) incorporation of phase transition and thermomechanical effects have been investigated (*dashed line* in Fig.9a). The conspicuous deviation of the wavelengths indicates the significance of, predominantly, the Marangoni flow. Nevertheless, the fluid movement-related effects influence more the final lattice displacement compared to the substantially smaller strain-generated movement. It is important to note, though, that hydrodynamics and lattice movements do not lead to suppression of ripples. This type of behaviour has also been observed in silicon upon irradiation with femtosecond pulses [20].

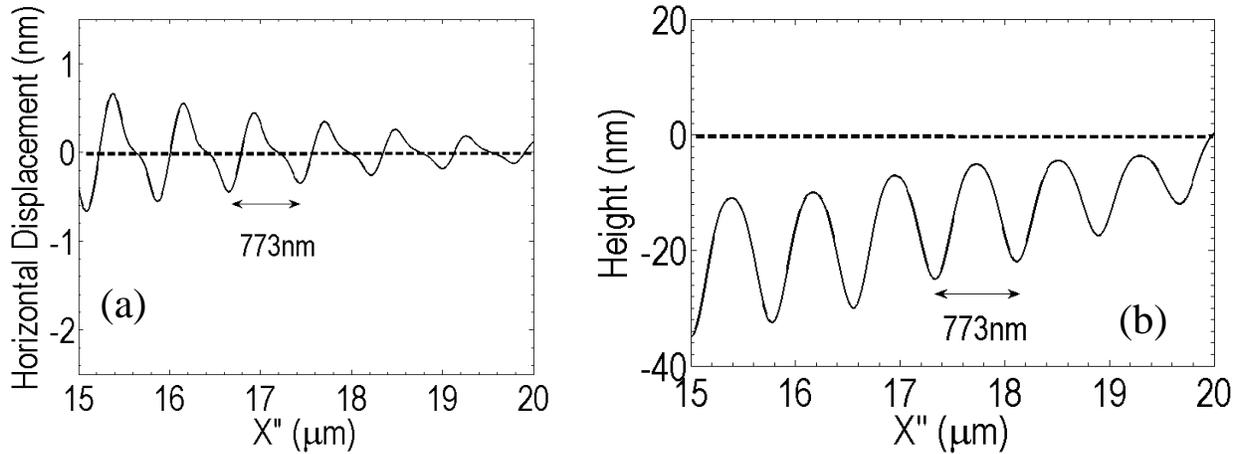

**Fig. 7** (a) Horizontal displacement and (b) Height (due to hydrodynamics) of the surface in the enlarged region ($X'' \in [15\mu m, 20\mu m]$), at $t=0.45$ns ($NP=2$).

Moreover, simulations indicate that irradiation with RP laser pulses results in larger ripple periodicity values, compared to the estimates derived using linearly polarized beams (Fig.9b). Firstly, in both types of polarization the ripple periodicity decreases with increasing number of pulses; as the number of pulses increases, the depth of the corrugated surface becomes larger due to mass displacement (as explained at the end of Section 2.6). Hence, the periodicity of the induced SP decreases (Fig.3) that eventually leads to a decreasing periodicity for the produced ripples [13]. On the other hand, the comparison of the predicted ripple periodicities for the two types of polarisation indicates that this deviation should be attributed to effects related to the heat distribution on the surface and relaxation processes. Specifically, as the lattice temperature reached on the surface of the material is higher for LP than RP (Fig.4), an enhanced heat diffusion to greater depths is produced; this effect in conjunction to the development of an increased temperature gradient on the surface and enhanced fluid vortex development eventually leads to hydrodynamical movement that leads to an increase of the ripple periodicity [16]. Hence, in contrast to the predicted (and experimentally observed) results in transition materials such as Ni with decreasing $g$ for increasing $T_e$ [35] and eventually higher $T_L$ for RP than that for LP, for noble materials (such as Ag) with the opposite electron-phonon coupling constant monotonicity, the periodicity values for RP follow the opposite trend.

Compared to other studies that focus on surface morphology with LP [16], it is evident that use of RP polarisation is associated with: (a) a different spatial profile of the beam that influences also the spatial distribution of energy deposition, absorption and eventually, the induced surface morphology, (b) a smaller maximum fluence peak that leads to a surface modification with smaller surface depth. However, the difference in the surface morphology induced by RP is not

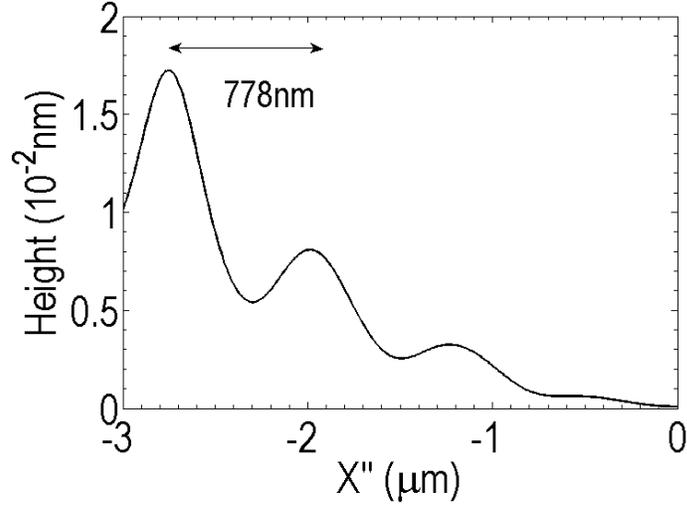

**Fig. 8.** Height of periodic structures (due to strain generation) of the surface close to $X''=0$, at $t=0.45$ns ($NP=2$).

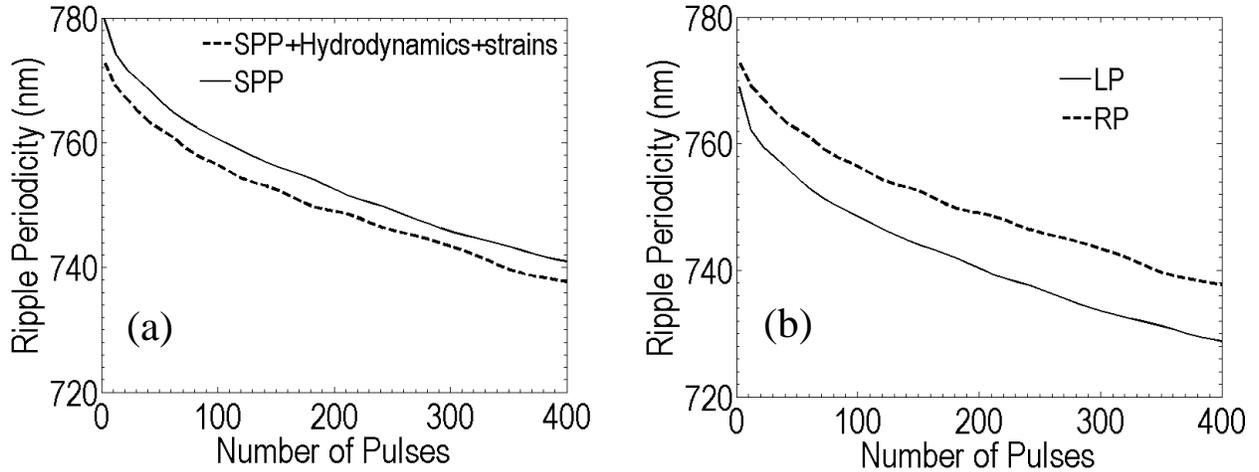

**Fig. 9** (a) Ripple periodicity dependence on number pulses with RP. Differences in the ripple periodicity are illustrated to highlight the significance of hydrodynamics and lattice displacements, (b) Ripple periodicity for RP and linearly polarized beams as a function of number of pulses. Simulation results correspond the average value of the periodicity inside the craters of the laser spot.

only related with the pronounced enhanced ripple periodicity or depth due to the dissimilar spatial profile of the deposited energy but also with a different orientation of the produced profile. More specifically, the ripple formation induced by RP (see Fig.10 for $NP=2$) are concentric periodic structures that have also been observed experimentally in previous works [17, 33, 35, 40, 41]. Indeed, a simulated two-dimensional pattern that is illustrated in Fig.10a shows the surface profile while the height distribution is shown in Fig.10b. On the other hand, the simulated profile predicted in previous studies suggests that different surface patterns are obtained by LP beams [16]. Although the same procedure towards simulating the induced surface modification can be followed for other values of *NP*, the images illustrated in this work attempt to highlight the transition from *NP*=1 (i.e. development of a crater) to *NP*=2 (development of ripples through the interference of the incident beam with a corrugated region). Hence, the rippled profile is repeated for *NP*>2. Regarding the significance of the periodic structures produced by RP beams, it has been shown that dynamic surface processing with this polarisation gives rise to formation of biomimetic structures [33]. This is achieved by exploiting the unique and versatile *angular* profile of femtosecond RP beams. The latter is one of the advantages of the use of RP laser beams as it enables formation of structures with more complex profiles than those produced with LP beams. Hence, laser processing with RP beams has a great potential to provide a plethora of complex structures and biomimetic surfaces and this justifies the need to elucidate further the underlying physical mechanisms.

A plausible question is related to whether the spatial profile of fluence of the RP beam is anticipated to influence the periodicity of the induced periodic structures. Firstly, simulations suggest that only one type of periodic structures is predicted (for the applied laser beam conditions): ripples with orientation perpendicular to the polarization of the RP and, a computed, very small deviation of the periodicity (<1nm inside each crater formed by RP) of the approximately 7 ripples formed inside the crater (in the region $X'' \in [15\mu m, 20\mu m]$ in Fig.7b). On the other hand, comparison of result in Fig.7,8 indicate that there is a small difference in the predicted values for ripples produced inside the crater (Fig.7) or where thermomechanical effects dominate (Fig.8). The discrepancy is due to the variation of the fluence value but also the magnitude of forces that lead to surface modification (i.e. surface tension or strain generation). Experimental studies indicate that variance of fluence and/or number of pulses should lead to a combination of structures (ripples and/or grooves) with periodicities that are smaller or larger than the wavelength of the laser beam [17, 40, 67] and different/same orientation. However, ripple periodicity shows a very weak dependence on fluence changes even in conditions were ablation occurs [33, 40, 41, 78] and therefore the spatial profile of fluence is not expected to produce substantially different ripple periodicities across the same spot. From a theoretical point of view, this is explained by the discussion in Section E; more specifically, the peak fluence used in the simulations is not sufficient to lead to corrugated areas of large depth (which would be followed by a substantially spatially variable energy absorption). Hence, the exclusion of ablation and further increase of the corrugation heights does not allow to observe periodical structures of notably variable frequency.

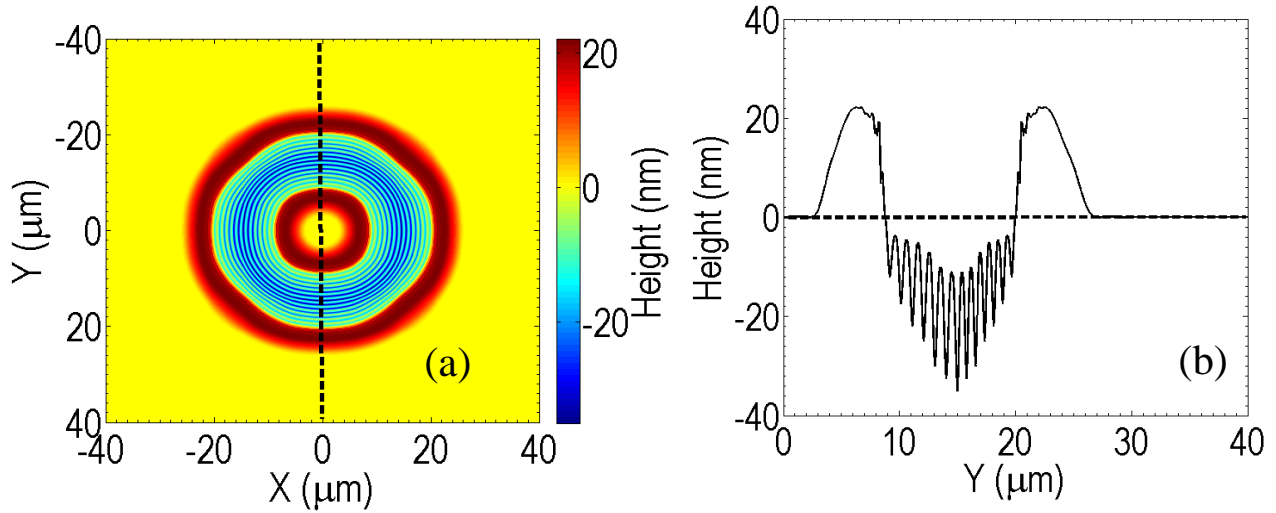

**Fig. 10** (Color Online) (a) Quadrant of a surface profile produced after irradiation with RP (*NP*=2), (b) Height profile along *dashed* line and *Y*>0 in (a).

To summarise, the simulation results produced in this work indicate that the model which describes energy absorption, electron excitation, heat transfer, relaxation processes enriched with a combination of phase change and thermomechanical response of the material offers a well-defined and consistent approach to explain the surface modification process upon irradiation with femtosecond RP pulses. As shown above, there are two distinct regions (around and away from the centre) where phase change and elastoplastic effects, respectively, dominate the underlying surface modification physical process. Hence, the aforementioned modelling approach allows the prediction of both the frequency of the induced periodic structures and the size (i.e. height) of the modified surfaces. It was shown that the spatial form of the RP does not only lead to structures of different periodicity compared to those induced if LP beams are used (Fig.9) but also produce structures of (spatially) variable height (Fig.7-8) and spatial distribution (Fig.10b). This is a significant prediction (which was verified experimentally for Ni in a previous work [33, 35] or semiconductors [40, 41]) as it demonstrates the capability of controlling characteristics of morphological features with the use of different polarization states. It should be mentioned, though, that a systematic experimental validation of the numerical results is required to confirm the accuracy of the proposed model for irradiation of noble metals with small electron-phonon coupling constant.

## IV   Conclusions

In conclusion, we have performed numerical simulations to provide a detailed analysis of the underlying mechanisms that lead to surface modification upon irradiation of Ag with RP fs laser pulses. The investigation shows the significant influence of the incident beam polarization on both the morphological profile as well as the size of the produced periodic structures. Despite the emphasis on morphological effects on noble metals, the study can also be extended to other types of solids, including dielectrics, semiconductors or polymers. Furthermore, while the emphasis of the work was focused on the morphological effects to RP beams, a similar approach can be followed for other types of CVB, namely, the azimuthally polarized beams. The ability to control the size of the morphological changes via modulating the beam polarization is expected to provide novel types of surface structures with significant advantages. Therefore, the proposed theoretical approach can be regarded as a complementary tool towards a systematic emerging laser based fabrication technique which can be exploited for expanding the breadth and novelty of potential applications [32, 79].

**ACKNOWLEDGEMENTS**

The authors acknowledge *LiNaBiofluid* (funded by EU's H2020 framework programme for research and innovation under Grant Agreement No. 665337) and N*anoscience Foundries and Fine Analysis* (NFFA)–Europe H2020-INFRAIA-2014-2015 (Grant agreement No 654360) projects for financial support.